\documentclass{iopart}
\usepackage{iopams}
\usepackage{graphicx}
\makeatletter

\providecommand{\LyX}{L\kern-.1667em\lower.25em\hbox{Y}\kern-.125emX\@}

\makeatother

\begin{document}

\title[RG for quantum $1/2$ spins on the pyroclore lattice]{Real-space renormalization group method for quantum $1/2$ spins on the pyrochlore lattice}
\author{Angel J.~Garcia-Adeva}
\address{Dpto.~Fisica Aplicada I, E.T.S.~Ingenieria de Bilbao, Universidad del Pais Vasco (UPV/EHU), Alda.~Urquijo s/n, 48013 Bilbao, Spain}
\address{Dpto.~Fisica Aplicada I, E.U.~Ingenieria de Vitoria, Universidad del Pais Vasco (UPV/EHU), C/~Nieves Cano 12, 01006 Vitoria, Spain} 
\ead{angel.garcia-adeva@ehu.es}

\begin{abstract}
A simple phenomenological real-space renormalization group method for quantum Heisenberg spins with nearest and next nearest neighbour interactions on a pyrochlore lattice is presented. Assuming a scaling law for the order parameter of two clusters of different sizes a set of coupled equations that gives the fixed points of the renormalization group transformation and, thus, the critical temperatures and ordered phases of the system is found. The particular case of spins $\frac{1}{2}$ is studied in detail. Furthermore, to simplify the mathematical details, from all the possible phases arising from the renormalization group transformation, only those phases in which the magnetic lattice is commensurate with a subdivision of the crystal lattice into four interlocked fcc sublattices are considered. These correspond to a quantum spin liquid, ferromagnetic order, or non-collinear order in which the total magnetic moment of a tetrahedral unit is zero. The corresponding phase diagram is constructed and the differences with respect to the classical model are analysed. It is found that this method reproduces fairly well the phase diagram of the pyrochlore lattice under the aforementioned constrains.
\end{abstract}

\section{Introduction}
Geometrically frustrated antiferromagnets exhibit a wealth of novel phases and a diverse set of new physical phenomena at low temperature \cite{Lacroix2011,Gardner2010,Balents2010,Moessner2006}. For this reason, there has been a renewed interest in these systems in the last decade. Some of the most studied materials are the spin ice compounds, in which the magnetic moments can be regarded as classical spins residing on the corners of a pyrochlore lattice with strong easy-axis anisotropy. Geometrical frustration in these systems leads to a classical spin liquid that displays Coulombic correlations and emergent magnetic monopole excitations that have been extensively studied \cite{Castelnovo2008,Bramwell2009,Savary2013a}. Even richer, however, could be the physics of the strong quantum effects coupled to strong geometrical frustration of Heisenberg spins on the pyrochlore lattice. It offers a route to unconventional phases in magnetism such as quantum spin liquids \cite{Savary2012a} or chiral order \cite{Choi2013} just to cite a couple of examples. 

The pyrochlore lattice is probably the lattice that exhibits the larger degree of frustration in three dimensions. It is a non-Bravais lattice consisting of four interlocked fcc lattices. Also, it can be seen as a lattice of corner sharing tetrahedra. A good starting point for a theoretical understanding of geometrical frustration in this lattice is the classical Heisenberg model with only isotropic nearest neighbour (NN) interactions. In this model, the non-trivial degeneracy of the ground state precludes any long range order and the system remains paramagnetic down to zero temperature \cite{Reimers1991,Garcia-Adeva2002b}. Since due to frustration the NN interaction does not fix the energy scale of the problem, however, any perturbation, no matter how small, can select one of the many degenerate states and the system can order. These perturbations can be anything including next nearest neighbour (NNN) Heisenberg interactions, dilution, or quantum fluctuations, to cite some examples. Quantum fluctuations in particular, as stressed above, seem to yield to exotic phases even in relatively simple models. For this reason, it is important to develop theoretical methods that provide a correct and simple description of the isotropic quantum Heisenberg model on the pyrochlore lattice while, at the same time, they can be easily extended to include additional, more complex interactions.

The present work reports on a real space renormalization group method that provides a good starting point for investigating theoretically the properties of quantum spins on the pyrochlore lattice: the so-called effective field renormalization group (EFRG) method. This work is an extension of earlier work on classical spins on the pyrochore lattice  \cite{Garcia-Adeva2002b} and the quantum version of the constant coupling method \cite{Garcia-Adeva2001d}. The latter, in particular, allows one to calculate the magnetic susceptibility for this lattice in a simple albeit quite exact manner. The present method, however, further allows to investigate the critical behaviour of this lattice (and several related ones, such as the kagome lattice) and to explore its phase diagram under the effect of various perturbations beyond the isotropic quantum Heisenberg interactions. The goal of this manuscript is to provide a comprehensive description of both  the foundations of the method in terms of the scaling hypothesis \cite{GOLDENFELD1994} and its application to calculating the phase diagram of quantum spins $\frac{1}{2}$ with both NN and NNN interactions on the pyrochlore lattice. It must be stressed that the present work does not try to give a full picture of the complete phase diagram of this system. Rather, our intention is to illustrate its use and the approximations involved in the analytical results by means of a relatively simple example. For this reason, constraints have been placed on the possible states of the system predicted by this method. In particular, only the reduced subset of magnetic phases that are commensurable with a division of the pyrochlore lattice in four interlocked fcc lattices has been studied. With this constraint, the set of equations for the order parameters reduces to four coupled equations that lead to three possible phases: a quantum spin liquid or cooperative paramagnet, a ferromagnetic long-range ordered state, or a non-collinear state in which the total magnetic moment of any tetrahedral unit is zero. The corresponding transition temperatures (if any) between the different phases can be easily calculated in this simple case and the corresponding phase diagrams for both ferromagnetic and antiferromagnetic NN interactions can be built. Although it is clear that this simplified phase diagram  is far from being complete, it provides a reasonable description of the reduced phases subset and a good starting point to implement further perturbations.

\section{The EFRG method for the pyrochlore lattice}
The central idea of any phenomenological renormalization group method \cite{PLASCAK1999} is based on the finite-size scaling hypothesis \cite{GOLDENFELD1994}. According to this, for any magnetic system defined by a Hamiltonian $\mathcal{H}=\mathcal{H}(K,H)$, where $K$ is the set of coupling constants and $H$ is an external magnetic field, there is a relationship between the values of any given thermodynamic quantity $P$ of two clusters of the physical system of different finite sizes $L$ and $L'$, respectively
\begin{equation}
\label{finite.scaling2}
\frac{P_{L'}(K',H')}{{L'}^{\phi }}=\frac{P_{L}(K,H)}{L^{\phi }}.
\end{equation}
In this expression, $\phi=\sigma/\nu$ is the anomalous dimension ($\sigma$ is the critical exponent for the quantity $P$ and $\nu$ is the correlation length critical exponent). $K'$ and $H'=l^{y_H}H$ are the rescaled coupling constants and magnetic field, respectively, under the scale transformation of the system given by the scale factor $l=L/L'$ ($y_H$ is the magnetic critical exponent). Depending on the particular quantity chosen for $P$ (the correlation length, the magnetization, \ldots), different phenomenological RG approaches result. 

In particular, one of the most successful approaches for spin systems on a lattice is the so called Effective Field RG (EFRG) method. This corresponds to applying relation \eref{finite.scaling2} to the order parameter of two spin clusters of different sizes (the sizes in this case are set by the number of spins in the clusters) calculated by means of the Callen-Suzuki identity \cite{CALLEN1963,SUZUKI1965}
\begin{equation}
\label{Callen1}
\left\langle O_{p}\right\rangle =\left\langle \frac{\Tr_{p}O_{p}\rme^{\mathcal{H}_{p}}}{\Tr_{p}\rme^{\mathcal{H}_{p}}}\right\rangle _{\mathcal{H}},
\end{equation}
for a cluster of size $L=p$, where the partial trace is taken over the set of $p$ spin variables specified
by the finite size cluster Hamiltonian $\mathcal{H}_{p}$; $O_{p}$
is the corresponding order parameter, and $\left\langle \ldots \right\rangle _{\mathcal{H}}$
indicates the usual canonical thermal average over the ensemble defined by the
complete Hamiltonian $\mathcal{H}$. The rest of spins outside the cluster that are not included in the partial trace in \eref{Callen1} are substituted by unknown values of the order parameter that act as a symmetry breaking field (SBF), $b$. The order parameter of the finite cluster can be computed for $b\ll 1$ and $H\ll 1$. An equation of the form
\begin{equation}
m_{p}(K,b,H)=\left\langle O_{p}\right\rangle =f_{p}(K)\, b+g_{p}(K)\, H,
\end{equation}
is obtained. Repeating the same calculation for a cluster of different size $L'=p'$ and assuming that both magnetizations are related by the scaling relation \eref{finite.scaling2} near criticality,  the following equation is obtained
\begin{equation}
\label{recursion.general}
f_{p'}(K')\, b'+g_{p'}(K')\, H'=l^{-\phi}\, f_{p}(K)\, b+l^{-\phi}\, g_{p}(K)\, H.
\end{equation}
 As $b$ and $b'$ are, in some sense, magnetizations, they are also assumed to satisfy \eref{finite.scaling2}, so one has the following relation for $H=0$
\begin{equation}
\label{recursion.thermal}
f_{p'}(K')=f_{p}(K),
\end{equation}
from which the fixed points of this RG transformation and, thus the critical points, are found by solving this equation for $K'=K=K_{c}$.

As for any real space formulation of the RG method, the only source of inaccuracy in these relations is in the finite sizes of the clusters. However, it has been found that phenomenological RG approaches are applicable to rather small systems and, sometimes, they are even more accurate than other more commonly used RG approaches, for clusters of adequate size \cite{PLASCAK1999}.

\section{Critical behaviour of the antiferromagnetic pyrochlore lattice}
The spins on a pyrochlore lattice occupy the corners of a 3D arrangement of corner sharing tetrahedra. As pointed out in other works \cite{Garcia-Adeva2002b,Garcia-Adeva2001c}, the ground state of the classical pyrochlore lattice (the one given by the condition that the total spin of any tetrahedron is zero) cannot be described by a single order parameter and, thus, the simple description of the EFRG given above is not actually valid for this lattice. Instead, it is necessary to break the non-Bravais pyrochlore lattice into four interlocked FCC (Bravais) lattices so that both NN and NNN spins belong to different sublattices. In terms of these sublattices, the Hamiltonian of the pyrochlore lattice with both NN and NNN interactions in the presence of a uniform magnetic field can be put in the form
\begin{equation}
\mathcal{H}=K_{1}\sum _{\alpha \neq \beta }\sum _{\left\langle i,m\right\rangle }\vec{s}_{i\alpha }\cdot \vec{s}_{j\beta }+K_{2}\sum _{\alpha \neq \beta }\sum _{\left\langle \left\langle i,m\right\rangle \right\rangle }\vec{s}_{i\alpha }\cdot \vec{s}_{j\beta }+\vec{H}_{0}\cdot \sum _{\alpha ,i}\vec{s}_{i\alpha },
\end{equation}
where $\vec{s}_{i\alpha }$ are the quantum Heisenberg spins of length $\left|\vec{s}_{i\alpha }\right|=s_0$. The $\alpha$ index labels the sublattice to which the considered spin belongs (and takes the values $\alpha =A,B,C,D$ for the pyrochlore lattice),
whereas the $i$ index labels the spins in a given sublattice (or, equivalently, the tetrahedron to which it belongs). $\left\langle \ldots \right\rangle$ represents the sum over NN pairs, whereas $\left\langle \left\langle \ldots \right\rangle \right\rangle$ stands for the sum over NNN pairs. $ K_{1}=\frac{J_{1}}{T}$ ($ K>0$
for ferromagnetic interactions and $K<0$ for antiferromagnetic ones) is the dimensionless NN exchange interaction and $K_{2}=\frac{J_{2}}{T}$ is the dimensionless NNN exchange interaction. Finally, $\vec{H}_{0}=\frac{\vec{h}_{0}}{T},$ with $\vec{h}_{0}$ the applied magnetic field.

In order to apply the EFRG method to this system, let us evaluate the order parameter of a spin belonging to a given sublattice for two clusters of sizes $p=1$ and $p'=4$.

\subsection*{The one-spin cluster}
The Hamiltonian of the one-spin cluster that belongs to sublattice $\alpha$ and the $i$-th tetrahedron can be cast in the form
\begin{equation}
\mathcal{H}_{1,i\alpha }=\vec{s}_{i\alpha }\cdot \vec{\xi }_{1,i\alpha },
\end{equation}
 where the subindex $1,i\alpha$ means that this Hamiltonian corresponds to a cluster of 1 spin that belongs to sublattice $\alpha $ and tetrahedron $i$, and $\vec{\xi }_{1,i\alpha }$ is the corresponding SBF given by 
\begin{equation}
\label{sbf.1spin}
\vec{\xi }_{1,i\alpha }=\vec{H}_{0}+K_{1}\sum _{\beta \neq \alpha }\sum _{j\in \mathrm{NN}}\vec{s}_{j\beta }+K_{2}\sum _{\beta \neq \alpha }\sum _{j\in \mathrm{NNN}}\vec{s}_{j\beta },
\end{equation}
where the sums over $j$ and $\beta$ in the second and third terms are restricted to NN and NNN on sublattice $\beta$, respectively (there are 2 NN and 4 NNN on each sublattice for the pyrochlore lattice). The order parameter for this cluster can be trivially evaluated using \eref{Callen1} \cite{SMART1966} and its value is
\begin{equation}
\vec{m}_{1\alpha }\simeq s_0(s_0+1)\frac{\left\langle \vec{\xi }_{1,i\alpha }\right\rangle _{\mathcal{H}}}{3}
\end{equation}
up to first order in the SBF.

\subsection*{The four-spin cluster}
The Hamiltonian of a cluster of 4 spins that belong to different sublattices and to the $i$-th tetrahedron in the presence of the symmetry breaking fields created by both the NN and NNN spins outside the cluster can be cast in the form
\begin{equation}
\label{hamiltonian.cluster}
\mathcal{H}_{4,i}=K_1\sum _{\left\langle \alpha ,\beta \right\rangle }\vec{s}_{i\alpha }\cdot \vec{s}_{i\beta }+\sum _{\alpha }\vec{s}_{i\alpha }\cdot \vec{\xi }_{4,i\alpha },
\end{equation}
 where the SBF acting on spin $i\alpha$ is given by
\begin{equation}
\label{sbf.4spin}
\vec{\xi}_{4,i\alpha}=\vec{H}_{0}+K_{1}\sum _{\beta \neq \alpha }\sum_{j\neq i\in \mathrm{NN}}\vec{s}_{j\beta }+K_{2}\sum _{\beta \neq \alpha }\sum _{j\in \mathrm{NNN}}\vec{s}_{j\beta }.
\end{equation}
In this equation, the sum in the second term of the right hand side member contains NN outside the $i$-th tetrahedron only.

The order parameter for this cluster given by \eref{Callen1} can be put in the alternative form
\begin{equation}
\label{partition.4spin}
\vec{m}_{4\alpha }=\left\langle \nabla _{\vec{\xi }_{4,i\alpha }}\ln Z_{4,i}\right\rangle _{{\mathcal{H}}},
\end{equation}
in terms of the partition function of the $i$-th cluster given by
\begin{equation}
Z_{4,i}=\Tr\, \rme^{\mathcal{H}_{4,i}},
\end{equation}
where the trace is calculated over the spins in the $i$-th tetrahedron. Of course, this partition function cannot be evaluated as a closed analytical expression in the general case but, in order to obtain an expression for the order parameter to first order in the SBF, it is enough to obtain a series expansion up to quadratic terms in this SBF. This is carried out in detail in \ref{ApA}. The result can be expressed in terms of the eigenstates  of the Hamiltonian on an isolated tetrahedral cluster, $\left| \gamma ,S,S_{z}\right\rangle $, where $S$ is the total spin momentum of the cluster, $S_{z}$ is its $z$ component, and $\gamma $ distinguishes between different eigenstates associated to the same eigenvalue and their corresponding eigenvalues
\begin{equation}
\label{eigenvalue.H0}
E_{S}=\frac{K_1}{2}S(S+1).
\end{equation}
The resulting expression is
\begin{equation}
\label{partition.series}
Z_{4,i}(\vec{\xi}_{4,i\alpha})=z_{0}+z_{2}(\vec{\xi}_{4,i\alpha}),
\end{equation}
where
\begin{equation}
\label{z0.term}
z_{0}(K_1)=\sum _{S=0}^{4s_0}g(S)(2S+1)\rme^{\frac{K_1}{2}S(S+1)},
\end{equation}
with $g(S)$ the degeneracy associated to a given value of $S$, which can be calculated by using Van Vleck's formula \cite{VLECK1959,Garcia-Adeva2001f}, and
\begin{eqnarray}
\label{z2.term}
\fl z_{2}(\vec{\xi}_{4,i\alpha})=\frac{1}{2}\sum _{\alpha , \beta }\left( \xi _{4,i\alpha }^{x}\xi _{4,i\beta }^{x}+\xi _{4,i\alpha }^{y}\xi _{4,i\beta }^{y}\right) \sum _{S,S'}\exp{\left(\frac{K_1}{2}S(S+1)\right)}\, G_{SS'}\nonumber\\
\times\sum _{\gamma ,\gamma '=1}^{g(S)}\sum _{S_{z},S'_{z}=-S}^{S}\left\langle \gamma ,S,S_{z}\right| \mathbf{s}_{i\alpha }^{+}\left| \gamma ',S',S'_{z}\right\rangle \left\langle \gamma ',S',S'_{z}\right| \mathbf{s}_{i\beta }^{-}\left| \gamma ,S,S_{z}\right\rangle\nonumber\\ 
\fl+\sum _{\alpha , \beta } \xi _{4,i\alpha }^{z}\xi _{4,i\beta }^{z} \sum _{S,S'}\exp{\left(\frac{K_1}{2}S(S+1)\right)}\, G_{SS'}\nonumber\\
\times\sum _{\gamma ,\gamma '=1}^{g(S)}\sum _{S_{z},S'_{z}=-S}^{S}\left\langle \gamma ,S,S_{z}\right| \mathbf{s}_{i\alpha }^{z}\left| \gamma ',S',S'_{z}\right\rangle \left\langle \gamma ',S',S'_{z}\right| \mathbf{s}_{i\beta }^{z}\left| \gamma ,S,S_{z}\right\rangle.
\end{eqnarray}
In this expression, $\xi_{4,i\alpha}^x$, $\xi_{4,i\alpha}^y$, and $\xi_{4,i\alpha}^z$ are the $x$, $y$, and $z$ components of the SBF $\mathbf{s}_{i\alpha}^\pm$ are the standard raising and lowering operators acting \emph{on the individual spin states}, and 
\begin{equation}
G_{SS'}=\frac{1}{E_{S}-E_{S'}}+\frac{\rme^{-(E_{S}-E_{S'})}-1}{(E_{S}-E_{S'})^{2}}.
\end{equation}

Equations \eref{partition.series}, \eref{z0.term}, and \eref{z2.term} provide an analytical expression of the partition function of the four-spin cluster up to second order in the SBF from which the order parameter can be evaluated in a simple way for arbitrary (even different among themselves) values of the individual spins of the cluster. They constitute one of the central results of this paper. Obviously, it is very complicated to evaluate $z_2$ in a closed form for arbitrary values of the individual spins. For this reason, in the following, we will limit the discussion to the simpler, albeit rather interesting, $s_0=\frac{1}{2}$ case.

\subsection*{The spin-\protect$\frac{1}{2}\protect$ case}
For $s_0=\frac{1}{2}$ the basis of eigenstates of the isolated tetrahedral cluster contains 16 elements only, which renders the problem manageable. In this basis, it is quite straightforward to evaluate \eref{z0.term} by taking into account \eref{eigenvalue.H0} and the fact that for four $\frac{1}{2}$ spins, the total spin momentum of the unit takes the values $S=0, 1,2$ with degeneracies $2,3,1$ and the corresponding eigenvalues of the Hamiltonian of the cluster are $0,K_1,3K_1$, respectively. Therefore, 
\begin{equation}
z_0(K_1)=2 + 9 \rme^{K_1} + 5 \rme^{3 K_1}.
\end{equation}

The calculation of the $z_2$ quadratic term, on the other hand, it is quite cumbersome and it is described in detail in \ref{ApB}. The result coming out from such a lengthy calculation is that the partition function, up to quadratic terms, can be put in the form
\begin{equation}
Z_{4,i}=z_{0}(K_1)+F(K_1)\sum _{\alpha }\xi _{4,i\alpha }^{2}+G(K_1)\sum _{\alpha \neq \beta }\vec{\xi }_{4,i\alpha }\cdot \vec{\xi }_{4,i\beta },
\end{equation}
where
\numparts
\begin{eqnarray}
F(K_1)&=\frac{-8+3(1+3K_1)\rme^{K_1}+5(1+K_1)\rme^{3K_1}}{16K_1}, \\
G(K_1)&=\frac{8+3(-1+K_1)\rme^{K_1}+5(-1+3K_1)\rme^{3K_1}}{48K_1}.
\end{eqnarray}
\endnumparts
It is interesting to notice that this form of the partition function is the most general one compatible with the symmetry of the isotropic interaction between the spins.

The order parameter of the cluster is easily calculated by using \eref{partition.4spin} up to linear terms in the SBF
\begin{equation}
\vec{m}_{4\alpha }=\frac{2F(K_1)}{z_{0}(K_1)}\left\langle \vec{\xi }_{4,i\alpha }\right\rangle +\frac{2G(K_1)}{z_{0}(K_1)}\sum _{\beta \neq \alpha }\left\langle \vec{\xi }_{4,i\beta }\right\rangle .
\end{equation}
On the other hand, for the 1-spin cluster
\begin{equation}
\vec{m}_{1\alpha }=s_{0}(s_{0}+1)\frac{\left\langle \vec{\xi }_{1\alpha }\right\rangle }{3}=\frac{\left\langle \vec{\xi }_{1\alpha }\right\rangle }{4},
\end{equation}
where the $\mathcal{H}$ subscripts in the thermal averages have been dropped for the sake of clarity. These two equations together with the scaling relation \eref{recursion.general} constitute the central expressions that allow one to investigate the critical behavior of the present system. However, even in the simple limit of small SBFs, it is still quite complicated to solve the recursion relation in order to obtain analytical expressions for the quantities of interest. The main reason for this is that the SBF contains spins outside the tetrahedral unit. To see this in more detail, let us consider in detail the form of the thermal averages of the SBF given by \eref{sbf.1spin} and \eref{sbf.4spin} for both the one- and four-spin clusters, respectively. The main difference between these two expressions lies on the fact that the SBF for the one-spin cluster contains a sum over the six NN of the $i\alpha$ spin, whereas the SBF for the four-spin cluster contains a sum over only the three NN of the $i\alpha$ spin that were not included in the cluster itself. The sum over NNN contains the same spins in both cases. The recursion relation for a given spin on the cluster under consideration does not fix, in principle, the order parameter for spins outside that cluster. For those, their corresponding recursion relation should be applied. In this way, a set of as many self-consistent equations as spins in the system would be obtained for the as many order parameters as spins in the system. This is, of course, a general feature of all real-space method and there are two ways to circumvent this: on the one hand, there is always the possibility of using a momentum-space formulation of the phenomenological RG method, the other is to supply some empirical information on the (expected) ground states of the system. The first approach would require a complete reformulation of the EFRG from scratch. In other contexts, this approach usually yields expansions of the quantities of interest that are better behaved in the mathematical sense at the expense of a more involved mathematical formulation. The second approach is better suited for the real-space formulation of these phenomenological RG methods and it is the one used in the present work to illustrate its usefulness. It is important to stress that even with its limitations, the present method is far superior than the standard mean-field theory mainly for two reasons: firstly, it implements certain correlations between the spins in an exact way; secondly, the SBF is not directly the magnetization, as in standard mean field theory, but rather, it is fixed by a self-consistent condition given by the scaling relation. This leads to more accurate expressions for certain critical properties (see for example \cite{Garcia-Adeva2001c}, where these facts are discussed in more detail). 

The simplest ordered state that can be investigated in the framework of the interlocked sublattice division of the pyrochlore lattice is the state in which spins belonging to the same sublattice possess the same value of the order parameter. Of course, this is an oversimplified description, as it leaves out some well-known possible ground states such as incommensurate ones \cite{Choi2013} or hexagonal antiferromagnetic loops \cite{Lee2002}. However, this simple choice is a good example to illustrate how the present method works and it still provides a rich enough picture of the magnetic ground state of this lattice. Besides, it can later be perturbed (by adding more sublattices, for example, as discussed in \cite{SMART1966}) to obtain other possible ground states. With this assumption, one can express the thermal average of the SBFs in terms of the internal fields on a given sublattice $\vec{h}'_{\alpha }=K_1\left\langle \vec{s}_{i\alpha }\right\rangle_\mathcal{H}$ as 
\begin{equation}
\left\langle\vec{\xi }_{1\alpha }\right\rangle_\mathcal{H}=\left\langle\vec{\xi }_{1,i\alpha }\right\rangle_\mathcal{H}=\vec{h}_{0}+2(1+2\lambda)\sum_{\beta \neq \alpha }\vec{h}'_{\beta },
\end{equation}
 and\begin{equation}
\left\langle\vec{\xi }_{4\alpha }\right\rangle_\mathcal{H}=\left\langle\vec{\xi }_{4,i\alpha }\right\rangle_\mathcal{H}=\vec{h}_{0}+(1+4\lambda)\sum _{\beta \neq \alpha }\vec{h}'_{\beta },
\end{equation}
where $\lambda=\frac{J_2}{J_1}$ and it has been taken into account the fact that for the pyrochlore lattice there are 2 NN and 4 NNN on each sublattice. The recursion relation \eref{recursion.general} now reads
\begin{equation}
\label{system.magnet}
\Xi(K_1,\lambda)\vec{h}'_{\alpha }+\Theta(K_1,\lambda)\sum _{\beta \neq \alpha }\vec{h}'_{\beta }=\Lambda(K_1,\lambda)\vec{h}_{0},
\end{equation}
where
\numparts
\begin{eqnarray}
\Xi(K_1,\lambda)      &= 6(1+4\lambda)\frac{G(K_1)}{z_0(K_1)}\nonumber\\
					  &=(1+4 \lambda)\frac{8+3(K_1-1)\rme^{K_1} +5(3 K_1-1)\rme^{3 K_1}}{8K_1\left(2+9 \rme^{K_1}+5 \rme^{3K_1}\right)},\\
\Theta(K_1,\lambda)   &= 2(1+4\lambda)\left[\frac{F(K_1)}{z_0(K_1)}+2\frac{G(K_1)}{z_0(K_1)}\right]-\frac{1+2\lambda}{2}\nonumber\\
 &=\frac{-8 (3 K_1+1)+ (3-75 K_1)\rme^{K_1}+(5-15 K_1)\rme^{3 K_1}}{24K_1 \left(2+9 \rme^{K_1}+5 \rme^{3K_1}\right)}\nonumber\\
&+4\lambda\frac{-4 (3 K_1+2)+(3-21 K_1)\rme^{K_1} +5 (3 K_1+1)\rme^{3 K_1} }{24 K_1\left(2+9 \rme^{K_1}+5 \rme^{3K_1}\right)}\\
\Lambda(K_1,\lambda)  &= \frac{1}{4}-\left[\frac{2F(K_1)}{z_0(K_1)}+\frac{6G(K_1)}{z_0(K_1)}\right]=\frac{2+3 \rme^{K_1}-5 \rme^{3 K_1}}{4\left(2+9 \rme^{K_1}+5 \rme^{3K_1}\right)}.
\end{eqnarray}
\endnumparts
The solution of this set of four equations with four unknowns provides all the possible behaviours of the system under the restrictions posed above.

\subsection*{Phase diagram}
For $\vec{h}_{0}\neq \vec{0}$, the above system of equations has the unique solution
\begin{equation}
\vec{h}'_{\alpha }=\frac{\Lambda(K_1,\lambda)}{\Xi(K_1),\lambda+3\Theta(K_1,\lambda)}\vec{h}_{0} \qquad (\alpha=A,B,C,D).
\end{equation}
This is the paramagnetic phase of the system in which all the spins try to align parallel to the external applied field. In fact, taking into account that \cite{Garcia-Adeva2001d}
\begin{equation}
\frac{2F(K_1)}{z_{0}(K_1)}+\frac{6G(K_1)}{z_{0}(K_1)}=\frac{\left\langle S^{2}\right\rangle_{ \mathcal{H}_0} }{3p}=\frac{3\rme^{K_1}+5\rme^{3K_1}}{4+18\rme^{K_1}+10\rme^{3K_1}}
\end{equation}
with $p=4$ and introducing the function 
\begin{equation}
\epsilon (K_1)=\frac{\left\langle S^{2}\right\rangle_{ \mathcal{H}_0} }{ps_{0}(s_{0}+1)}-1= \frac{-2-3\rme^{K_1}+5\rme^{3K_1}}{2+9\rme^{K_1}+5\rme^{3K_1}},
\end{equation}
a connection with the results reported in \cite{Garcia-Adeva2001d} and \cite{Garcia-Adeva2001e} in the framework of the generalized constant coupling method can be furnished for $\lambda=0$ and $s_0=1/2$. Indeed, the internal fields can be put in the form 
\begin{equation}
\vec{h}'_{\alpha }=-\frac{\epsilon(K_1)}{3(\epsilon (K_1)-1)+12\lambda\epsilon(K_1)}\vec{h}_0.
\end{equation}
Substituting back in the expression of $\left\langle\xi_{1\alpha}\right\rangle_\mathcal{H}$ (or $\left\langle\xi_{4\alpha}\right\rangle_\mathcal{H}$) and subsequently in $\vec{m}_{1\alpha}$ (or $\vec{m}_{4,\alpha}$)
one can very easily calculate the susceptibility of the system to be
\begin{equation}
\chi(T)=\frac{1}{4T}\frac{1+\epsilon (K_1)}{1-(1+4\lambda)\epsilon(K_1)}.
\end{equation}
In particular, for NN interactions only ($\lambda=0$), one gets the expression
\begin{equation}
\chi(T)=\frac{1}{4T}\frac{1+\epsilon (K_1)}{1-\epsilon(K_1)},
\end{equation}
in complete agreement with the results reported in \cite{Garcia-Adeva2001e} and \cite{Garcia-Adeva2001d}.

In the absence of external applied field, $\vec{h}_0=0$, the set of equations \eref{system.magnet} has two possible solutions: on the one hand, for $\Xi(K_1,\lambda)+3\Theta(K_1,\lambda)=0$ or, equivalently, 
\begin{equation}
\label{condition.ferro}
\epsilon(K_1)=\frac{1}{(1+4\lambda)},
\end{equation}
the system is in a ferromagnetic state in which
\begin{equation}
\vec{m}_{iA}=\vec{m}_{iB}=\vec{m}_{iC}=\vec{m}_{iD} \qquad (i=1,2,\ldots,N).
\end{equation}
On the other hand, for $\Xi(K_1,\lambda)=\Theta(K_1,\lambda)$ or, equivalently,
\begin{equation}
\label{condition.nc}
\frac{F(K_1)}{z_0(K_1)}-\frac{G(K_1)}{z_0(K_1)}=\frac{-8+(3+6K_1)\rme^{K_1}+5\rme^{3K_1}}{12K_1(2+9\rme^{K_1}+5\rme^{3K_1})}=\frac{1}{4}\frac{1+2\lambda}{1+4\lambda},
\end{equation}
the system is in a non-collinear state (sometimes also called the $\vec{q}=\mathbf{0}$ state \cite{Reimers1991}) given by
\begin{equation}
\sum_\alpha \vec{m}_{i\alpha}=\vec{0},\qquad (i=1,2,\ldots,N),
\end{equation}
that is, the ground state in which the total spin of the tetrahedral unit is zero.

\begin{figure}
\includegraphics[width=\textwidth]{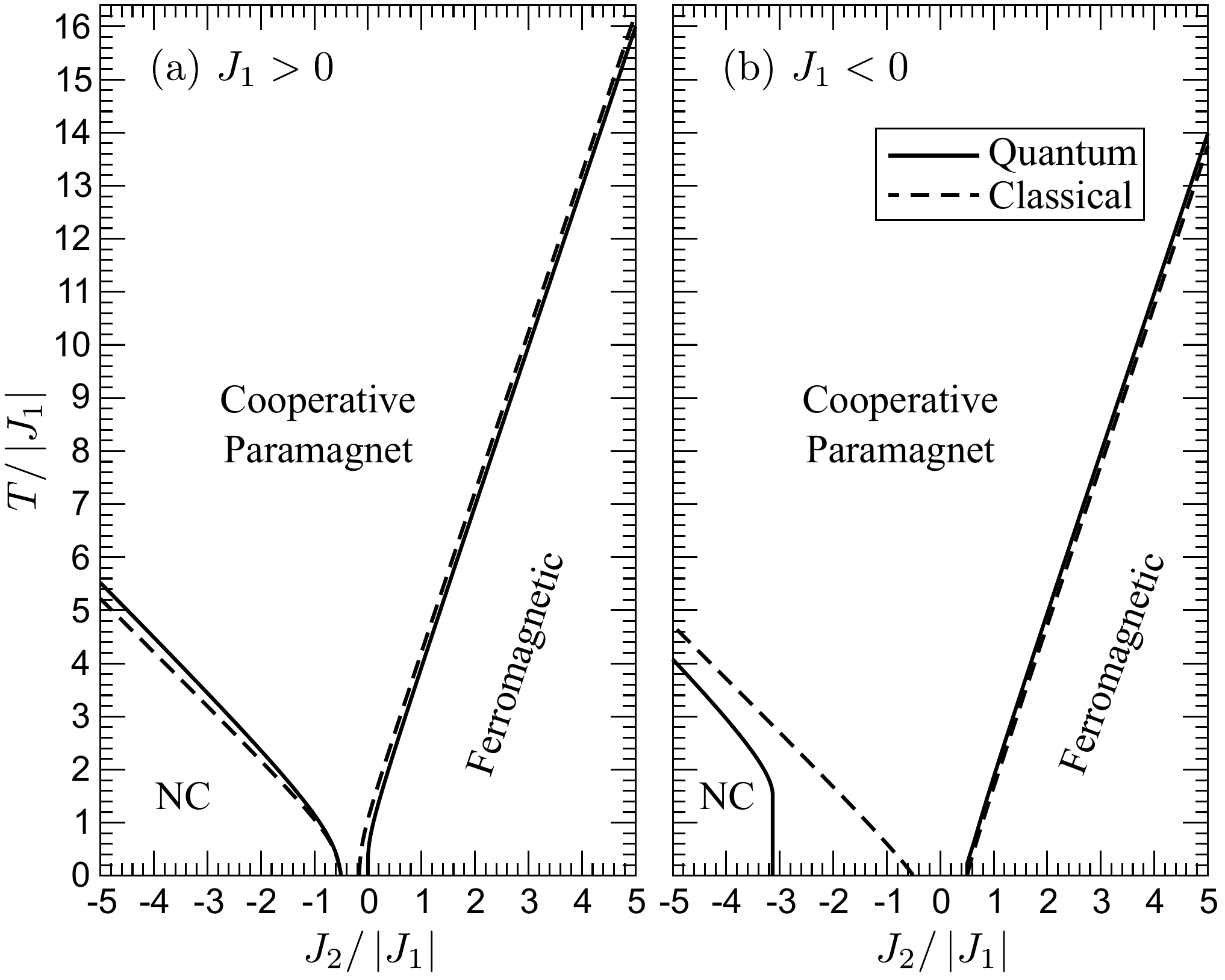}
\caption{\label{phase.diagram}$T-J_2$ phase diagram of the quantum Heisenberg spin $\frac{1}{2}$ on the pyrochlore lattice for (a) ferromagnetic NN interactions and (b) antiferromagnetic NN interactions. NC stands for the non-collinear phase. Discontinuous lines depict the corresponding phase boundaries for classical spins.}
\end{figure}
Conditions \eref{condition.ferro} and \eref{condition.nc} allows one to calculate the temperature at which the ferromagnetic or non-collinear states, respectively, set in as a function of the NN and NNN coupling constants ratio, i.e., they provide the phase diagram of the system with the constrains posed above. This is depicted in figures \ref{phase.diagram}a and \ref{phase.diagram}b for ferromagnetic and antiferromagnetic NN interactions, respectively. For clarity, the analogous phase diagram for classical Heisenberg spins (of length $s_0=\frac{1}{2}$ for the sake of comparison) has been also depicted. The quantum phase diagram exhibits a number of very interesting features for both ferro- and antiferromagnetic NN interactions.

On the ferromagnetic side ($J_1>0$), the first striking feature is that quantum fluctuations preclude the formation of the ferromagnetic state when only NN interactions are present and the system remains in a quantum spin liquid (or cooperative paramagnet) state down to $T=0$ K. This is in contrast with the results of this same RG method for classical spins, which predicts a transition to a ferromagnetic state at $T_c\approx1.15J_1$ \cite{Garcia-Adeva2002b}, a value relatively close to the one found by Champion \etal ($T_c\approx1.4J_1$) \cite{CHAMPION2002} in the limit of negligible single ion anisotropy by using the Monte Carlo method. Indeed, the classical ferromagnetic state is so robust that it persists even in the presence of antiferromagnetic NNN interactions down to $\lambda=-\frac{1}{6}$. For quantum spins, however, it is necessary the presence of NNN ferromagnetic interactions (no matter how small) to stabilize ferromagnetic long range order. It is very likely that the ferromagnetic state sets in for quantum Heisenberg spins $\frac{1}{2}$ in higher dimensions and also in three dimensions for spins larger than $\frac{1}{2}$ even in the absence of NNN interactions. Although outside the scope of this work, it would be very interesting to check these facts in the framework of the present model. In the range $-0.5<\lambda<0$ the system remains in a QSL state down to $T=0$ K. Below $\lambda=-0.5$, however, NNN antiferromagnetic interactions stabilize the NC state at a finite temperature.

For antiferromagnetic NN interactions ($J_1<0$), frustration precludes magnetic ordering in an even larger range. For $-3.132<-\lambda<0.5$ (notice that, since $J_1<0$, $-\lambda=J_2/\left|J_1\right|$) the system remains in a QSL state down to $T=0$ K. For ferromagnetic NNN interactions above $-\lambda=0.5$, the system reaches a ferromagnetic ordered state at finite temperature for values very close to those found for classical spins. When the NNN interactions are antiferromagnetic, the situation is quite different from the one for classical spins. In the later, there is a transition to a non-collinear phase at finite temperature for $\lambda<-0.5$. In the quantum case, however, the system abruptly goes from no transition at all to a transition from the QSL state to the NC state at $T_c=-1.52J_1$. For stronger antiferromagnetic NNN interactions, the value of the critical temperature asymptotically goes towards the classical value.

\section{Conclusions}
The present work reports on a real-space phenomenological renormalization group method for investigating the critical behaviour of quantum Heisenberg spins on a pyrochlore lattice with nearest and next nearest neighbour interactions. The central idea of the method is based on the finite-scaling hypothesis. According to this, the order parameters of two clusters of different sizes are related by a recursion relation from which the critical temperatures and critical exponents can be calculated. In the present context, two clusters consisting of one and four (a tetrahedral unit of the lattice) spins have been used in order to illustrate the method. We have also made use of the fact that the pyrochlore lattice is a non-Bravais lattice consisting of four interlocked fcc lattices. The calculation of the order parameter of the one-spin cluster is trivial. However, the corresponding calculation for the tetrahedral unit is a very complex problem. We have developed an iterative expansion for this quantity in powers of the symmetry breaking fields due to the rest of spins outside the tetrahedral cluster that interact with the ones in the cluster. Since obtaining a general result for arbitrary spin values is very difficult, the expansion has been applied to the case of spins of length $\frac{1}{2}$. For the particular geometry of the pyrochlore lattice, the scaling relation results into an infinite set of self-consistent coupled equations for the thermal averages of the symmetry breaking fields that fix the order parameters of the system and allows one to study its critical behaviour. In order to gain any analytical insight on this behaviour without resorting to numerical solutions, the problem has been further simplified by restricting the set of possible solutions to those that fulfil the condition that spins into a given fcc sublattice posses the same value of the order parameter. With this constraint, the recursion relation results in four linear coupled equations that predict that the system can be found in one of three possible phases in the absence of applied magnetic field: a quantum spin liquid (or cooperative paramagnet), a ferromagnetic state, or a non-collinear state in which the total magnetization of a given tetrahedral cluster is zero. The corresponding transition temperatures between these phases can be also calculated and the corresponding phase diagram constructed. One of the most remarkable features of this phase diagram is that for NN ferromagnetic interactions only, quantum fluctuations destroy the ferromagnetic order in three dimensions, in contrast with both analytical and numerical results for classical spins on this lattice. Also, on the antiferromagnetic NN interaction sector --where frustration is expected to occur-- the method correctly predicts that no long-range order is stabilized at any finite temperature when only NN interactions are present. Moreover, even in the presence of  antiferromagnetic NNN interactions, it is necessary that these are sufficiently strong, above 3 times the NN ones, in order to be able to stabilize a non-collinear ordered state at finite temperature. Since, for real systems, the strength of NNN interactions is typically $\approx 0.1$ times the strength of the NN ones, it would be undoubtedly very useful to include other interactions in the present framework to compare its predictions against real systems.

Of course, the present method is not exempt of limitations and it is important to point them out while, at the same time, suggesting possible workarounds. The first limitation of any real-space method and its main source of inaccuracy is the finite size of the clusters: the smaller the clusters, the larger the error in the determination of the critical quantities of the system. Actually, in the present manuscript, we selected what it is possibly the worst case scenario: the two smallest clusters that take into account the  non-Bravais character of the pyrochlore lattice. Still, it is quite remarkable that the physical picture obtained with these sizes is very reasonable. This, in fact, seems to be a quite general feature of phenomenological renormalization group methods \cite{PLASCAK1999}. Unfortunately, the amount of work required to increase the size of the clusters is not commensurable with the refinement obtained in the critical quantities. What is even worse, in any real-space method, there is no way to estimate \textit{a priori} the error made for given cluster sizes. The only way to overcome this difficulty and to obtain a controlled expansion is to switch over to Fourier space. However, what is gained in mathematical rigour is overcome by the mathematical complexity associated to working in this alternative formulation. In any case, devising a $k$-space formulation of the present phenomenological RG method is also a desirable route that we intend to explore, since it is the only way in which, for example, incommensurate long-range ordered phases can be studied. Another complication of any real-space method is the fact that in order to be able to obtain more general states than the ones studied in this paper it is necessary to further divide the lattice into more sublattices. For example, to allow for states in which the NN and NNN in the same fcc sublattice posses different spin orientations, it is necessary to further divide each fcc lattice into four additional sublattices, resulting in a set of 16 coupled equations which, even though more complicated than the particular case analyzed in this work, are still manageable and lead to a much richer phase diagram. In any case, as mentioned above, it is also important to stress that this method, even with its limitations,  is far superior than standard mean-field theory. This is so mainly for two reasons: on the one hand, it implements correlations between the spins in a tetrahedral cluster in an exact way and these are, precisely, the correlations that play the most important role in a large part of the phase diagram of the pyrochlore lattice (in particular, in the spin liquid phase); on the other hand, the SBF produced by the spins outside the cluster is not directly the magnetization, as it is usually assumed in standard mean field theory, but rather, it is fixed by a self-consistent condition given by the scaling relation. These two facts lead to more accurate expressions for many quantities of the system. In particular, the predicted susceptibility in the paramagnetic regime is almost exact, as reported in \cite{Garcia-Adeva2001d} and \cite{Garcia-Adeva2001e}. 

In conclusion, we think it is fair to say that the quantum EFRG method provides a reliable and relatively simple theoretical tool to study the critical behaviour of quantum spins on the pyrochlore (and other geometrically frustrated) lattices. Even with the stringent constraints posed in the present manuscript to simplify the calculations and being able to obtain analytical approximations of the quantities of interest, the phase diagram obtained in this framework is rich and it seems qualitatively reasonable. We hope that other researchers will find it useful to investigate the exotic magnetic behaviour of this lattice at low temperatures.

\ack I would like to thanks the Spanish MICIIN for financial support under the I+D+i programme (grant FIS2009-14293).
\appendix
\section{Evaluation of the partition function of a tetrahedral spin cluster}\label{ApA}
In this Appendix, a detailed calculation of the partition function of a tetrahedral spin cluster is presented. The main difficulty of this calculation lies on the fact that, in contrast with classical spins, in order to make the expansion in powers of the SBF, it is necessary to take into account the non-commutativity of the Heisenberg and Zeeman terms in \eref{hamiltonian.cluster}.

Let us recast \eref{hamiltonian.cluster} in the following form
\begin{equation}
\mathcal{H}_{4,i}=\mathcal{H}_{0}+\mathcal{H}',
\end{equation}
where
\begin{equation}
\label{H0.hamiltonian}
\mathcal{H}_{0}=K_1\sum _{\left\langle \alpha ,\beta \right\rangle }\vec{s}_{1\alpha }\cdot \vec{s}_{1\beta }
\end{equation}
 and\begin{equation}
\label{zeeman.term}
\mathcal{H}'=\sum _{\alpha }\vec{s}_{i\alpha }\cdot \vec{\xi }_{4,i\alpha }.
\end{equation}
Next we construct the auxiliary operator
\begin{equation}
\mathcal{F}(\lambda )=\rme^{-\lambda \mathcal{H}_{0}}\rme^{\lambda \mathcal{H}_{4,i}}.
\end{equation}
 In terms of this function, the partition function of the cluster
can be put as
\begin{equation}
\label{partition.F}
Z_{4,i}=\Tr\left( \rme^{\mathcal{H}_{0}}\mathcal{F}(1)\right).
\end{equation}
Calculating a series expansion for the partition function is totally equivalent to calculating a series expansion for the $F(\lambda )$ operator. To do this, let us consider the elements of this
operator in an orthonormal discrete basis of $\mathcal{H}_{0}$, $\left\{ \left| n\right\rangle \right\} $,
such that\begin{equation}
F_{nm}(\lambda )=\left\langle n\right| \mathcal{F}(\lambda )\left| m\right\rangle =\rme^{-\lambda E_{n}}\left\langle n\right| \rme^{\lambda \mathcal{H}_{4,i}}\left| m\right\rangle ,
\end{equation}
 where $E_{n}=\left\langle n\right| \mathcal{H}_{0}\left| n\right\rangle $
is the eigenvalue of $\mathcal{H}_{0}$ for eigenstate $\left| n\right\rangle $.
Differentiating once\begin{equation}
\frac{\partial F_{nm}(\lambda )}{\partial \lambda }=\left\langle n\right| \rme^{-\lambda \mathcal{H}_{0}}\, \mathcal{H}'\, \rme^{\lambda \mathcal{H}_{0}}\, \mathcal{F}(\lambda )\left| m\right\rangle ,
\end{equation}
which can be put as the integral equation
\begin{equation}
\label{integral.equation}
F_{nm}(\lambda )=\delta _{nm}+\int _{0}^{\lambda }\left\langle n\right| \rme^{-\mu\mathcal{H}_{0}}\, \mathcal{H}'\, \rme^{\mu\mathcal{H}_{0}}\, \mathcal{F}(\mu)\left| m\right\rangle \, \rmd\mu,
\end{equation}
where we have taken into account the boundary condition $F_{nm}(0)=\delta _{nm}$, with $\delta_{nm}$ the Kronecker delta. Obviously, the zeroth order solution to this equation is
\begin{equation}
F_{nm}^{(0)}(\lambda )=\delta _{nm}.
\end{equation}
Substituting this solution in \eref{integral.equation} we obtain the first order approximation
\begin{eqnarray}
\fl F_{nm}^{(1)}(\lambda )=\delta _{nm}+\int _{0}^{\lambda }\left\langle n\right| \rme^{\mu\mathcal{H}_{0}}\, \mathcal{H}'\, \rme^{\mu\mathcal{H}_{0}}\left| m\right\rangle \, \rmd\mu=\delta _{nm}+\left\langle n\right| \mathcal{H}'\left| m\right\rangle \int _{0}^{\lambda }\rme^{-\mu(E_{n}-E_{m})}\, \rmd\mu\nonumber\\
=\delta _{nm}+\frac{1-\rme^{-\lambda (E_{n}-E_{m})}}{E_{n}-E_{m}} H'_{nm},
\end{eqnarray}
where $H'_{nm}=\left\langle n \right| \mathcal{H}' \left|m\right\rangle$. Substituting again in \eref{integral.equation} we arrive at the desired order of approximation for $F_{nm}(\lambda )$.
\begin{eqnarray}
\fl F_{nm}^{(2)}(\lambda )=\delta _{nm}+\int _{0}^{\lambda }\left\langle n\right| \rme^{-\mu\mathcal{H}_{0}}\, \mathcal{H}'\, \rme^{\mu\mathcal{H}_{0}}\, \mathcal{F}^{(1)}(\mu)\left| m\right\rangle \, \rmd\mu\nonumber\\
=\delta _{nm}+\sum _{l}\int _{0}^{\lambda }\left\langle n\right| \rme^{-\mu\mathcal{H}_{0}}\, \mathcal{H}'\, \rme^{\mu\mathcal{H}_{0}}\left| l\right\rangle \left\langle l\right| \mathcal{F}^{(1)}(\mu)\left| m\right\rangle\rmd\mu\nonumber \\ 
=\delta _{nm}+F_{nm}^{(1)}(\lambda )+\sum _{l}\left\langle n\right| \mathcal {H}'\left| l\right\rangle \left\langle l\right| \mathcal {H}'\left| m\right\rangle \int ^{\lambda }_{0}\frac{1-\rme^{-\mu (E_{n}-E_{m})}}{E_{n}-E_{m}}\, \rmd\mu\nonumber\\ 
=\delta _{nm}+\frac{1-\rme^{-\lambda (E_{n}-E_{m})}}{E_{n}-E_{m}}H'_{nm}\nonumber\\
+\sum _{l}\left[ \frac{1-\rme^{-\lambda (E_{n}-E_{l})}}{(E_{n}-E_{l}) (E_{l}-E_{m})} - \frac{1-\rme^{-\lambda (E_{n}-E_{m})}}{(E_{n}-E_{m})(E_{l}-E_{m})}\right] H'_{nl}H'_{lj}.
\end{eqnarray}
In principle, this iterative process can be continued indefinitely. For our purposes, however, this is the required order of the expansion to calculate the order parameter up to linear terms in the SBF. Taking into account \eref{partition.F}, the partition function of the cluster takes a very simple form up quadratic terms in the SBF
\begin{equation}
\label{partition.series.ap}
Z_{4,i}=\sum _{n}\rme^{E_{n}}\, F_{nn}(1)=z_{0}+z_{1}+z_{2},
\end{equation}
where
\begin{equation}
\label{z0.ap}
z_{0}=\sum _{n}\rme^{E_{n}},
\end{equation}
\begin{equation}
\label{z1.ap}
z_{1}=\sum _{i}\rme^{E_{n}}\, H'_{nn},
\end{equation}
and
\begin{equation}
\label{z2.ap}
z_{2}=\sum_{n,m}\rme^{E_{n}}\, G_{nm}\, {H'}_{nm}^{2},
\end{equation}
with
\begin{equation}
G_{nm}=\frac{1}{E_{n}-E_{m}}+\frac{\rme^{-(E_{n}-E_{m})}-1}{(E_{n}-E_{m})^{2}}.
\end{equation}
It is interesting to notice that $G_{nn}=\frac{1}{2}$.

Let us now apply \eref{partition.series} to the Hamiltonian of the cluster given by \eref{hamiltonian.cluster}. The eigenstates of $\mathcal{H}_{0}$ are of the form $\left| \gamma ,S,S_{z}\right\rangle $, where $S$ is the total spin momentum of the cluster, $S_{z}$ is its $z$ component, and $\gamma $ distinguishes between different eigenstates associated to the same eigenvalue. The corresponding eigenvalues of $\mathcal{H}_{0}$ are of the form
\begin{equation}
E_{S}=\frac{K_1}{2}S(S+1).
\end{equation}
Therefore, \eref{z0.ap} can be put in the form
\begin{equation}
\label{z0.term.ap}
z_{0}=\sum _{S}g(S)(2S+1)\rme^{\frac{K_1}{2}S(S+1)},
\end{equation}
 with $g(S)$ the degeneracy associated to a given value of $S$, which can be calculated by using Van Vleck's formula \cite{VLECK1959,Garcia-Adeva2001f}. On the other hand, the $z_{1}$ term is zero because the eigenstates of $\mathcal{H}_{0}$ have defined parity. Finally, the $z_{2}$
term can be put in the form
\begin{equation}
\label{z2.dos.ap}
z_{2}=\sum _{S,S'}\rme^{E_{S}}\, G_{SS'}\sum _{\gamma ,\gamma '=1}^{g(S)}\sum _{S_{z},S'_{z}=-S}^{S}\left| \left\langle \gamma ,S,S_{z}\right| \mathcal{H}'\left| \gamma ',S',S'_{z}\right\rangle \right| ^{2}.
\end{equation}
 By using \eref{zeeman.term} and introducing the spin raising and lowering operators for the individual spins of the cluster
 \begin{equation}
\mathbf{s}_{i\alpha }^{\pm }=\mathbf{s}_{i\alpha }^{x}\pm \rmi\, \mathbf{s}_{i\alpha }^{y},
\end{equation}
 the only non-zero terms in \eref{z2.dos.ap} are
\begin{eqnarray}
\label{z2.term.ap}
\fl z_{2}=\frac{1}{2}\sum _{\alpha , \beta }\left( \xi _{4,i\alpha }^{x}\xi _{4,i\beta }^{x}+\xi _{4,i\alpha }^{y}\xi _{4,i\beta }^{y}\right) \sum _{S,S'}\rme^{E_{S}}\, G_{SS'}\nonumber\\
\times\sum _{\gamma ,\gamma '=1}^{g(S)}\sum _{S_{z},S'_{z}=-S}^{S}\left\langle \gamma ,S,S_{z}\right| \mathbf{s}_{i\alpha }^{+}\left| \gamma ',S',S'_{z}\right\rangle \left\langle \gamma ',S',S'_{z}\right| \mathbf{s}_{i\beta }^{-}\left| \gamma ,S,S_{z}\right\rangle\nonumber\\ 
\fl+\sum _{\alpha , \beta } \xi _{4,i\alpha }^{z}\xi _{4,i\beta }^{z} \sum _{S,S'}\rme^{E_{S}}\, G_{SS'}\nonumber\\
\times\sum _{\gamma ,\gamma '=1}^{g(S)}\sum _{S_{z},S'_{z}=-S}^{S}\left\langle \gamma ,S,S_{z}\right| \mathbf{s}_{i\alpha }^{z}\left| \gamma ',S',S'_{z}\right\rangle \left\langle \gamma ',S',S'_{z}\right| \mathbf{s}_{i\beta }^{z}\left| \gamma ,S,S_{z}\right\rangle.
\end{eqnarray}

\section{Evaluation of $z_2(K_1)$ for $s_0=\frac{1}{2}$}
\label{ApB}
The evaluation of the quadratic term \eref{z2.term} is a cumbersome task even in the simplest $s_0=\frac{1}{2}$ case. The first thing one needs to do in order to carry out this calculation is to select an adequate basis in which to calculate the matrix elements appearing in \eref{z2.term}. This is the basis of the eigenvectors of the Hamiltonian of the $i$-th isolated tetrahedral cluster (given by \eref{H0.hamiltonian}), which can be expressed as linear combinations of the basis of the vector space spanned by the tensor product of the four individual spin states as follows
\begin{eqnarray}
\label{eigenstates}
\left| 1,0,0\right\rangle  & =  \case{1}{2}\left[ \left| ++--\right\rangle +\left|--++\right\rangle -\left|+-+-\right\rangle -\left| -+-+\right\rangle \right]\\
\left| 2,0,0\right\rangle  & = \case{1}{2\sqrt{3}}\left[ \left| ++--\right\rangle +\left|--++\right\rangle +\left|+-+-\right\rangle\right.\nonumber\\
&\left.+\left|-+-+\right\rangle-2\left|+--+\right\rangle-2\left|-++-\right\rangle\right]\\
\left| 1,1,1\right\rangle  & =  \case{1}{2}\left[\left|+++-\right\rangle -\left| ++-+\right\rangle - \left|+-++\right\rangle +\left|-+++\right\rangle\right] \\
\left| 2,1,1\right\rangle  & =  \case{1}{2}\left[\left|+++-\right\rangle +\left|++-+\right\rangle - \left|+-++\right\rangle -\left|-+++\right\rangle\right] \\
\left| 3,1,1\right\rangle  & =  \case{1}{2}\left[\left|+++-\right\rangle -\left|++-+\right\rangle + \left|+-++\right\rangle -\left|-+++\right\rangle\right] \\
\left| 1,1,0\right\rangle  & = \case{1}{\sqrt{2}}\left[ -\left| -++-\right\rangle +\left| +--+\right\rangle \right] \\
\left| 2,1,0\right\rangle  & = \case{1}{\sqrt{2}}\left[ -\left| ++--\right\rangle +\left| --++\right\rangle \right] \\
\left| 3,1,0\right\rangle  & =  \case{1}{\sqrt{2}}\left[ -\left| +-+-\right\rangle +\left| -+-+\right\rangle \right] \\
\left| 1,1,-1\right\rangle  & = \case{1}{2}\left[\left|---+\right\rangle -\left| --+-\right\rangle - \left|-+--\right\rangle +\left|+---\right\rangle\right] \\
\left| 2,1,-1\right\rangle  & =  \case{1}{2}\left[\left|---+\right\rangle +\left|--+-\right\rangle - \left|-+--\right\rangle -\left|+---\right\rangle\right] \\
\left| 3,1,-1\right\rangle  & = \case{1}{2}\left[\left|---+\right\rangle -\left|--+-\right\rangle + \left|-+--\right\rangle -\left|+---\right\rangle\right] \\
\left| 1,2,2\right\rangle  & =  \left| ++++\right\rangle \\
\left| 1,2,1\right\rangle  & =  \case{1}{2}\left[ \left| +++-\right\rangle +\left| ++-+\right\rangle +\left| +-++\right\rangle +\left| -+++\right\rangle \right] \\
\left| 1,2,0\right\rangle  & =  \case{1}{\sqrt{6}}\left[ \left| ++--\right\rangle +\left| +-+-\right\rangle +\left| -++-\right\rangle \right. \nonumber \\
&   +\left. \left| +--+\right\rangle +\left| -+-+\right\rangle +\left| --++\right\rangle \right]\\
\left| 1,2,-1\right\rangle  & =  \case{1}{2}\left[ \left| ---+\right\rangle +\left| --+-\right\rangle +\left| -+--\right\rangle +\left| +---\right\rangle \right]\\
\left| 1,2,-2\right\rangle  & =  \left| ----\right\rangle.
\end{eqnarray}
In these expressions, the '$+$' symbol indicates a spin state with $s_z=\case{1}{2}$, whereas the '$-$' symbol represents the opposite state. It is important to stress that the operators appearing in \eref{z2.term} are one particle operators, that is, they act on a single spin of the unit. In particular, for $s_0=\case{1}{2}$, they fulfill the following relationships
\begin{equation}
\label{algebra}
\begin{array}{lll}
\mathbf{s}^z_{i\alpha}\left|+\right\rangle_{i\alpha} = \case{1}{2}\left|+\right\rangle_{i\alpha} &\mathbf{s}^+_{i\alpha}\left|+\right\rangle_{i\alpha} = 0
&\mathbf{s}^-_{i\alpha}\left|+\right\rangle_{i\alpha} = \left|-\right\rangle_{i\alpha}\\
\mathbf{s}^z_{i\alpha}\left|-\right\rangle_{i\alpha} = -\case{1}{2}\left|-\right\rangle_{i\alpha} &\mathbf{s}^+_{i\alpha}\left|-\right\rangle_{i\alpha} = \left|+\right\rangle_{i\alpha}
&\mathbf{s}^-_{i\alpha}\left|-\right\rangle_{i\alpha} = 0.
\end{array}
\end{equation}
The easiest way to perform the sums in \eref{z2.term} is probably to obtain the matrix form of the single particle operators in the basis  \eref{eigenstates}. This is accomplished by calculating all elements of the form
$\left\langle \gamma ,S,S_{z}\right| s_{i\alpha }^{\mu}\left| \gamma ',S',S'_{z}\right\rangle$ for $\mu=z,+,-$. To quote a couple of examples, let us calculate
\begin{eqnarray}
\left\langle 1 ,1,0\right| \mathbf{s}_{iA}^{+}\left| 1,1,0\right\rangle=\case{1}{2}\left[ \left\langle+--+\right|\mathbf{s}^+_{iA}\left|+--+\right\rangle \right.\nonumber\\
\left.- \left\langle+--+\right|\mathbf{s}^+_{iA}\left|-++-\right\rangle - \left\langle-++-\right|\mathbf{s}^+_{iA}\left|+--+\right\rangle \right.\nonumber\\
\left. + \left\langle-++-\right|\mathbf{s}^+_{iA}\left|-++-\right\rangle \right]= -\case{1}{2}\left\langle+--+\right.\left|+++-\right\rangle\nonumber\\
 + \case{1}{2}\left\langle-++-\right|\left.+++-\right\rangle =0
\end{eqnarray}
and
\begin{eqnarray}
\left\langle 1 ,1,0\right| \mathbf{s}_{iA}^{+}\left| 1,1,-1\right\rangle=\case{1}{2\sqrt{2}}\left[ \left\langle+--+\right|\mathbf{s}^+_{iA}\left|---+\right\rangle \right.\nonumber\\
\left.- \left\langle+--+\right|\mathbf{s}^+_{iA}\left|--+-\right\rangle - \left\langle+--+\right|\mathbf{s}^+_{iA}\left|-+--\right\rangle \right.\nonumber\\
+ \left\langle+--+\right|\mathbf{s}^+_{iA}\left|+---\right\rangle - \left\langle-++-\right|\mathbf{s}^+_{iA}\left|---+\right\rangle\nonumber\\
+ \left\langle-++-\right|\mathbf{s}^+_{iA}\left|--+-\right\rangle + \left\langle-++-\right|\mathbf{s}^+_{iA}\left|-+--\right\rangle\nonumber\\
\left.-\left\langle-++-\right|\mathbf{s}^+_{iA}\left|+---\right\rangle \right]= \case{1}{2\sqrt{2}}\left[\left\langle+--+\right.\left|+--+\right\rangle\right.\nonumber\\
-\left\langle+--+\right.\left|+-+-\right\rangle-\left\langle+--+\right.\left|++--\right\rangle\nonumber\\
-\left\langle-++-\right.\left|+--+\right\rangle+\left\langle-++-\right.\left|+-+-\right\rangle\nonumber\\
\left.+\left\langle-++-\right.\left|++--\right\rangle\right]=\frac{1}{2\sqrt{2}}
\end{eqnarray}
By repeating this procedure for the 16 basis vectors, for the four sublattices, and for the rest of operators, the matrix form of the single particle operators are obtained. For example, the ones operating on the $A$ sublattice are
\begin{equation}
\fl s^z_{iA}=\begin{tiny}\left(
\begin{array}{c@{\hspace{2pt}}c@{\hspace{2pt}}c@{\hspace{2pt}}c@{\hspace{2pt}}c@{\hspace{2pt}}c@{\hspace{2pt}}c@{\hspace{2pt}}c@{\hspace{2pt}}c@{\hspace{2pt}}c@{\hspace{2pt}}c@{\hspace{2pt}}c@{\hspace{2pt}}c@{\hspace{2pt}}c@{\hspace{2pt}}c@{\hspace{2pt}}c@{\hspace{2pt}}} 0 & 0 & 0 & 0 & 0 & 0 & -\frac{1}{2 \sqrt{2}} & \frac{1}{2 \sqrt{2}} & 0 & 0 & 0 & 0 & 0 & 0 & 0 & 0 \\
 0 & 0 & 0 & 0 & 0 & -\frac{1}{\sqrt{6}} & -\frac{1}{2 \sqrt{6}} & -\frac{1}{2 \sqrt{6}} & 0 & 0 & 0 & 0 & 0
   & 0 & 0 & 0 \\
 0 & 0 & \frac{1}{4} & \frac{1}{4} & \frac{1}{4} & 0 & 0 & 0 & 0 & 0 & 0 & 0 & -\frac{1}{4} & 0 & 0 & 0 \\
 0 & 0 & \frac{1}{4} & \frac{1}{4} & -\frac{1}{4} & 0 & 0 & 0 & 0 & 0 & 0 & 0 & \frac{1}{4} & 0 & 0 & 0 \\
 0 & 0 & \frac{1}{4} & -\frac{1}{4} & \frac{1}{4} & 0 & 0 & 0 & 0 & 0 & 0 & 0 & \frac{1}{4} & 0 & 0 & 0 \\
 0 & -\frac{1}{\sqrt{6}} & 0 & 0 & 0 & 0 & 0 & 0 & 0 & 0 & 0 & 0 & 0 & \frac{1}{2 \sqrt{3}} & 0 & 0 \\
 -\frac{1}{2 \sqrt{2}} & -\frac{1}{2 \sqrt{6}} & 0 & 0 & 0 & 0 & 0 & 0 & 0 & 0 & 0 & 0 & 0 & -\frac{1}{2
   \sqrt{3}} & 0 & 0 \\
 \frac{1}{2 \sqrt{2}} & -\frac{1}{2 \sqrt{6}} & 0 & 0 & 0 & 0 & 0 & 0 & 0 & 0 & 0 & 0 & 0 & -\frac{1}{2
   \sqrt{3}} & 0 & 0 \\
 0 & 0 & 0 & 0 & 0 & 0 & 0 & 0 & -\frac{1}{4} & -\frac{1}{4} & -\frac{1}{4} & 0 & 0 & 0 & \frac{1}{4} & 0 \\
 0 & 0 & 0 & 0 & 0 & 0 & 0 & 0 & -\frac{1}{4} & -\frac{1}{4} & \frac{1}{4} & 0 & 0 & 0 & -\frac{1}{4} & 0 \\
 0 & 0 & 0 & 0 & 0 & 0 & 0 & 0 & -\frac{1}{4} & \frac{1}{4} & -\frac{1}{4} & 0 & 0 & 0 & -\frac{1}{4} & 0 \\
 0 & 0 & 0 & 0 & 0 & 0 & 0 & 0 & 0 & 0 & 0 & \frac{1}{2} & 0 & 0 & 0 & 0 \\
 0 & 0 & -\frac{1}{4} & \frac{1}{4} & \frac{1}{4} & 0 & 0 & 0 & 0 & 0 & 0 & 0 & \frac{1}{4} & 0 & 0 & 0 \\
 0 & 0 & 0 & 0 & 0 & \frac{1}{2 \sqrt{3}} & -\frac{1}{2 \sqrt{3}} & -\frac{1}{2 \sqrt{3}} & 0 & 0 & 0 & 0 &
   0 & 0 & 0 & 0 \\
 0 & 0 & 0 & 0 & 0 & 0 & 0 & 0 & \frac{1}{4} & -\frac{1}{4} & -\frac{1}{4} & 0 & 0 & 0 & -\frac{1}{4} & 0 \\
 0 & 0 & 0 & 0 & 0 & 0 & 0 & 0 & 0 & 0 & 0 & 0 & 0 & 0 & 0 & -\frac{1}{2}
\end{array}
\right)\end{tiny}
\end{equation}
and
\begin{equation}
\fl\mathbf{s}^+_{iA}=(\mathbf{s}^-_{iA})^\dagger=\begin{tiny}
\left(
\begin{array}{c@{\hspace{2pt}}c@{\hspace{2pt}}c@{\hspace{2pt}}c@{\hspace{2pt}}c@{\hspace{2pt}}c@{\hspace{2pt}}c@{\hspace{2pt}}c@{\hspace{2pt}}c@{\hspace{2pt}}c@{\hspace{2pt}}c@{\hspace{2pt}}c@{\hspace{2pt}}c@{\hspace{2pt}}c@{\hspace{2pt}}c@{\hspace{2pt}}c@{\hspace{2pt}}}
 0 & 0 & 0 & 0 & 0 & 0 & -\frac{1}{2 \sqrt{2}} & \frac{1}{2 \sqrt{2}} & 0 & 0 & 0 & 0 & 0 & 0 & 0 & 0 \\
 0 & 0 & 0 & 0 & 0 & -\frac{1}{\sqrt{6}} & -\frac{1}{2 \sqrt{6}} & -\frac{1}{2 \sqrt{6}} & 0 & 0 & 0 & 0 & 0 & 0 & 0 & 0 \\
 0 & 0 & \frac{1}{4} & \frac{1}{4} & \frac{1}{4} & 0 & 0 & 0 & 0 & 0 & 0 & 0 & -\frac{1}{4} & 0 & 0 & 0 \\
 0 & 0 & \frac{1}{4} & \frac{1}{4} & -\frac{1}{4} & 0 & 0 & 0 & 0 & 0 & 0 & 0 & \frac{1}{4} & 0 & 0 & 0 \\
 0 & 0 & \frac{1}{4} & -\frac{1}{4} & \frac{1}{4} & 0 & 0 & 0 & 0 & 0 & 0 & 0 & \frac{1}{4} & 0 & 0 & 0 \\
 0 & -\frac{1}{\sqrt{6}} & 0 & 0 & 0 & 0 & 0 & 0 & 0 & 0 & 0 & 0 & 0 & \frac{1}{2 \sqrt{3}} & 0 & 0 \\
 -\frac{1}{2 \sqrt{2}} & -\frac{1}{2 \sqrt{6}} & 0 & 0 & 0 & 0 & 0 & 0 & 0 & 0 & 0 & 0 & 0 & -\frac{1}{2 \sqrt{3}} & 0 & 0 \\
 \frac{1}{2 \sqrt{2}} & -\frac{1}{2 \sqrt{6}} & 0 & 0 & 0 & 0 & 0 & 0 & 0 & 0 & 0 & 0 & 0 & -\frac{1}{2 \sqrt{3}} & 0 & 0 \\
 0 & 0 & 0 & 0 & 0 & 0 & 0 & 0 & -\frac{1}{4} & -\frac{1}{4} & -\frac{1}{4} & 0 & 0 & 0 & \frac{1}{4} & 0 \\
 0 & 0 & 0 & 0 & 0 & 0 & 0 & 0 & -\frac{1}{4} & -\frac{1}{4} & \frac{1}{4} & 0 & 0 & 0 & -\frac{1}{4} & 0 \\
 0 & 0 & 0 & 0 & 0 & 0 & 0 & 0 & -\frac{1}{4} & \frac{1}{4} & -\frac{1}{4} & 0 & 0 & 0 & -\frac{1}{4} & 0 \\
 0 & 0 & 0 & 0 & 0 & 0 & 0 & 0 & 0 & 0 & 0 & \frac{1}{2} & 0 & 0 & 0 & 0 \\
 0 & 0 & -\frac{1}{4} & \frac{1}{4} & \frac{1}{4} & 0 & 0 & 0 & 0 & 0 & 0 & 0 & \frac{1}{4} & 0 & 0 & 0 \\
 0 & 0 & 0 & 0 & 0 & \frac{1}{2 \sqrt{3}} & -\frac{1}{2 \sqrt{3}} & -\frac{1}{2 \sqrt{3}} & 0 & 0 & 0 & 0 & 0 & 0 & 0 & 0 \\
 0 & 0 & 0 & 0 & 0 & 0 & 0 & 0 & \frac{1}{4} & -\frac{1}{4} & -\frac{1}{4} & 0 & 0 & 0 & -\frac{1}{4} & 0 \\
 0 & 0 & 0 & 0 & 0 & 0 & 0 & 0 & 0 & 0 & 0 & 0 & 0 & 0 & 0 & -\frac{1}{2}
\end{array}
\right)\end{tiny}
\end{equation}
On the other hand, the function $G_{SS'}$ is more concisely tabulated as
\begin{equation*}
\begin{tabular}{c|ccc}
$G_{SS'}$ & 0 & 1 & 2\\\hline
0 & $\frac{1}{2}$ & $\frac{-1-K_1+\rme^{K_1}}{K_1^2}$ & $\frac{-1-3K_1+\rme^{3K_1}}{9K_1^2}$ \\
1 & $\frac{1+(K_1-1)\rme^{K_1}}{K_1^2}$ & $\frac{\rme^{K_1}}{2}$ & $\frac{-(1+2K_1)\rme^{K_1}+\rme^{3K_1}}{4K_1^2}$ \\
2 & $\frac{1+(3K_1-1)\rme^{3K_1}}{9K_1^2}$ & $\frac{\rme^{K_1}+(-1+2K_1)\rme^{3K_1}}{4K_1}$ & $\frac{\rme^{3K_1}}{2}$ \\
\end{tabular}
\end{equation*}

Collecting all the above results, it is relatively easy to calculate the sums in \eref{z2.term}. After simplification, the result one obtains is 
\begin{eqnarray}
\fl F(K_1)&=\sum _{S,S'}\rme^{E_{S}}G_{SS'}\sum _{\gamma ,\gamma '}\sum _{S_{z},S'_{z}}\left\langle \gamma ,S,S_{z}\right| \mathbf{s}_{i\alpha }^{z}\left| \gamma ',S',S'_{z}\right\rangle \left\langle \gamma ',S',S'_{z}\right| \mathbf{s}_{i\alpha }^{z}\left| \gamma ,S,S_{z}\right\rangle\nonumber\\
\fl &=\frac{1}{2}\sum _{S,S'}\rme^{E_{S}}G_{SS'}\sum _{\gamma ,\gamma '}\sum _{S_{z},S'_{z}}\left\langle \gamma ,S,S_{z}\right| s_{i\alpha }^{+}\left| \gamma ',S',S'_{z}\right\rangle \left\langle \gamma ',S',S'_{z}\right| \mathbf{s}_{i\alpha }^{-}\left| \gamma ,S,S_{z}\right\rangle\nonumber\\
\fl &=\frac{-8+3(1+3K_1)\rme^{K_1}+5(1+K_1)\rme^{3K_1}}{16K_1}, \\
\fl G(K_1)&=\sum _{S,S'}\rme^{E_{S}}G_{SS'}\sum _{\gamma ,\gamma '}\sum _{S_{z},S'_{z}}\left\langle \gamma ,S,S_{z}\right| \mathbf{s}_{i\alpha }^{z}\left| \gamma ',S',S'_{z}\right\rangle \left\langle \gamma ',S',S'_{z}\right| \mathbf{s}_{i\beta }^{z}\left| \gamma ,S,S_{z}\right\rangle\nonumber\\
\fl &=\frac{1}{2}\sum _{S,S'}\rme^{E_{S}}G_{SS'}\sum _{\gamma ,\gamma '}\sum _{S_{+},S'_{z}}\left\langle \gamma ,S,S_{z}\right| \mathbf{s}_{i\alpha }^{z}\left| \gamma ',S',S'_{z}\right\rangle \left\langle \gamma ',S',S'_{z}\right| \mathbf{s}_{i\beta }^{-}\left| \gamma ,S,S_{z}\right\rangle\nonumber\\ 
\fl &=\frac{8+3(-1+K_1)\rme^{K_1}+5(-1+3K_1)\rme^{3K_1}}{48K_1}.
\end{eqnarray}
The quadratic term is then given by
\begin{equation}
z_2=F(K_1)\sum_\alpha \xi_{4,i\alpha}^2+G(K_1)\sum_{\alpha\neq\beta} \vec{\xi}_{4,i\alpha}\cdot\vec{\xi}_{4,i\beta}.
\end{equation}
\section*{References}
\bibliographystyle{unsrt}

\begin{thebibliography}{10}

\bibitem{Lacroix2011}
C.~Lacroix, P.~Mendels, and F.~Mila.
\newblock {\em Introduction to Frustrated Magnetism: Materials, Experiments,
  Theory}.
\newblock Springer Series in Solid-State Sciences. Springer, Berlin,
  Heidelberg, 2011.

\bibitem{Gardner2010}
J.~S. Gardner, M.~J.~P. Gingras, and J.~E. Greedan.
\newblock Magnetic pyrochlore oxides.
\newblock {\em Rev. Mod. Phys.}, 82:53--107, Jan 2010.

\bibitem{Balents2010}
L.~Balents.
\newblock Spin liquids in frustrated magnets.
\newblock {\em Nature}, 464:199--208, March 2010.

\bibitem{Moessner2006}
R.~Moessner and A.~P. Ramirez.
\newblock Geometrical frustration.
\newblock {\em Physics Today}, 59:24--29, Feb 2006.

\bibitem{Castelnovo2008}
C.~Castelnovo, R.~Moessner, and S.~L. Sondhi.
\newblock Magnetic monopoles in spin ice.
\newblock {\em Nature}, 451:42--45, January 2008.

\bibitem{Bramwell2009}
S.~T. Bramwell, S.~R. Giblin, S.~Calder, R.~Aldus, D.~Prabhakaran, and
  T.~Fennell.
\newblock Measurement of the charge and current of magnetic monopoles in spin
  ice.
\newblock {\em Nature}, 461:956--959, October 2009.

\bibitem{Savary2013a}
L.~{Savary} and L.~{Balents}.
\newblock Spin liquid regimes at nonzero temperature in quantum spin ice.
\newblock {\em Phys. Rev. B}, 87:205130, May 2013.

\bibitem{Savary2012a}
L.~Savary and L.~Balents.
\newblock Coulombic quantum liquids in spin-$1/2$ pyrochlores.
\newblock {\em Phys. Rev. Lett.}, 108:037202, Jan 2012.

\bibitem{Choi2013}
E.~Choi, G.-W. Chern, and N.~B. Perkins.
\newblock Chiral magnetism and helimagnons in a pyrochlore antiferromagnet.
\newblock {\em Phys. Rev. B}, 87:054418, Feb 2013.

\bibitem{Reimers1991}
J.~N. Reimers, A.~J. Berlinsky, and A.~C. Shi.
\newblock Mean-field approach to magnetic-ordering in highly frustrated
  pyrochlores.
\newblock {\em Physical Review B}, 43(1):865--878, Jan 1 1991.

\bibitem{Garcia-Adeva2002b}
A.~J. Garcia-Adeva and D.~L. Huber.
\newblock {Classical generalized constant-coupling method for geometrically
  frustrated magnets: Microscopic formulation and effect of perturbations
  beyond nearest-neighbor interactions}.
\newblock {\em {Physical Review B}}, {65}({18}):184418, {May 1} {2002}.

\bibitem{Garcia-Adeva2001d}
A.~J. Garcia-Adeva and D.~L. Huber.
\newblock {Quantum generalized constant-coupling model for geometrically
  frustrated antiferromagnets}.
\newblock {\em {Physical Review B}}, {63}({17}):174433, {May 1} {2001}.

\bibitem{GOLDENFELD1994}
N.~Goldenfeld.
\newblock {\em Lectures on Phase Transitions and the Renormalization Group}.
\newblock Addison-Wesley, 1994.

\bibitem{PLASCAK1999}
J.~A. Plascak, W.~Figueiredo, and B.~C.~S. Grandi.
\newblock Phenomenological renormalization group methods.
\newblock {\em Brazilian Journal of Physics}, 29(3):579, 1999.

\bibitem{CALLEN1963}
H.~B. Callen.
\newblock A note on green functions and the ising model.
\newblock {\em Phys. Lett.}, 4:161, 1963.

\bibitem{SUZUKI1965}
M.~Suzuki.
\newblock Generalized exact formulation for the correlations of the ising model
  and other classical systems.
\newblock {\em Phys. Lett.}, 19:267, 1965.

\bibitem{Garcia-Adeva2001c}
A.~J. Garcia-Adeva and D.~L. Huber.
\newblock {Critical behavior of two- and three-dimensional ferromagnetic and
  antiferromagnetic spin-ice systems using the effective-field renormalization
  group technique}.
\newblock {\em {Physical Review B}}, {64}({1}):014418, {Jul 1} {2001}.

\bibitem{SMART1966}
J.~S. Smart.
\newblock {\em Effective field theories of magnetism}.
\newblock Saunders, Philadelphia, 1966.

\bibitem{VLECK1959}
J.~H. van Vleck.
\newblock {\em The theory of electric and magnetic susceptibilities}.
\newblock Oxford, London, 1959.

\bibitem{Garcia-Adeva2001f}
A.~J. Garcia-Adeva and D.~L. Huber.
\newblock {Quantum tetrahedral mean field theory of the magnetic susceptibility
  for the pyrochlore lattice}.
\newblock {\em {Physical Review Letters}}, {85}({21}):{4598--4601}, {Nov 20}
  {2000}.

\bibitem{Lee2002}
S.-H. Lee, C.~Broholm, W.~Ratcliff, G.~Gasparovic, Q.~Huang, T.~H. Kim, and
  S.-W Cheong.
\newblock Emergent excitations in a geometrically frustrated magnet.
\newblock {\em Nature}, 418:856--858, Aug 2002.

\bibitem{Garcia-Adeva2001e}
A.~J. Garcia-Adeva and D.~L. Huber.
\newblock {Classical generalized constant coupling model for geometrically
  frustrated antiferromagnets}.
\newblock {\em {Physical Review B}}, {63}({14}):140404, {Apr 1} {2001}.

\bibitem{CHAMPION2002}
J.~D.~M. Champion, S.~T. Bramwell, P.~C.~W. Holdsworth, and M.~J. Harris.
\newblock Competition between exchange and anisotropy in a pyrochlore
  ferromagnet.
\newblock {\em Europhysics Letters}, 57(1):93--99, Jan 2002.

\end{thebibliography}

\end{document}